# Impact and Implementation of Reserved Lanes for Automated Driving on Signalized Urban Arterials

**Slobodan Gutesa, Ph.D.**
Transportation Engineer
Greenman-Pedersen, Inc, New York, NY 10018
Email: sgutesa@gpinet.com

**ABSTRACT**

An automated vehicle refers to a vehicle that can achieve a safe movement on a roadway facility without the influence of a human driver. With emerging trend of the connected vehicle concept over the past decade, numerous state-of-the-art applications focusing on automated vehicle-based intersection control have been proposed. The main purpose of this study is to estimate and evaluate impact of designated lanes for automated vehicles and recommend some viable lane configuration scenarios for signalized urban arterials. The automated driving was simulated in PTV Vissim using trajectory-driven control strategy. The concept evaluation through microsimulation reveals significant mobility improvements compared to operational scenario without lane reservation. Findings imply that for signalized corridors observed in this study, total travel time reductions are ranging from 5.1% to 19.4% depending on C/AV market penetration, and test-bed configuration parameters.

## INTRODUCTION

Operating signalized corridor with lane reservation strategy represent a complex and labor-intensive task [1]. One of the main purposes to introduce lane reservation concept was to promote and increase attractiveness of the public transit [2-3]. The concept of designated lanes for high occupancy vehicles (HOV) is another concept that introduced lane reservation aspect to improve overall mobility of roadway facilities [4-7]. Following the same conceptual framework, many researchers in the past decade investigated implementation and promotion of automated driving using lane reservation approach [7-9], however, not too many researchers offered implementation strategy that would help roadway operators estimate impact of lane reservation on overall signalized corridor mobility under imperfect market penetration of automated vehicles. In contrast to existing knowledge, this research investigates operational strategy for signalized corridor under different market penetration levels. It also provides some general insight into C/AV market penetration and traffic volume thresholds relevant for initiation of the lane reservation when corridor is operated under C/AV environment.

## METHODOLOGY

The concept of reserved lanes for automated driving investigated in this study, assumes the inclusion of such lanes into existing signalized corridor using overhead gantries. The reserved lanes allow automated vehicles to be segregated from the general traffic in order to eliminate the interaction between two vehicle groups. The lane reservation concept allows smooth integration of automated vehicles assuming road users are well familiar with similar lane assignment concepts such as High Occupancy Vehicle (HOV) lanes, high-occupancy toll (HOT) lanes, or eco-lanes. The efficiency of such lanes is evaluated under uncongested (i.e. LOS C) and congested (i.e. LOS

E) conditions. Comparison between mobility performance measures for the corridor with and without reserved lanes gives insight into the applicability of the concept under different traffic conditions.

## Simulation of Automated Vehicles in PTV Vissim

In this study, movement of automated vehicles is modeled using control strategy introduce in the previous research [10]. The methodology presented in mentioned research introduces control strategy for a signalized arterial with fixed-timing signals under imperfect market penetration of the connected vehicle technology. The methodology assumes the provision of vehicular parameters (i.e. speed and position) by utilizing connected and automated vehicle environment (C/AV). Information exchange and proactive adjustment of vehicles' trajectories is handled by the computer system, namely TOAD control agent. The TOAD control agent collects necessary vehicular and signal status information from equipped vehicles and traffic signal controllers to determine the optimal speed for every automated vehicle in the system, while allowing regular, unequipped vehicles, to maintain safe and uninterrupted movement along the corridor.

## Corridor Design for Automated Driving with Reserved Lanes

To adequately integrate reserved lanes into signalized arterial, the optimal lane group to be used are inner lanes (i.e. left-most lanes). The main reason for described lane assignment is that the left-most lane allows uninterrupted movement of automated vehicles in cases where unequipped vehicles are making right turns at intersections or access points of the corridor.

To apply such lane assignment strategy, the signalized corridor must be equipped with jughandle intersections as left-most lanes cannot be used for left turns. Therefore, all vehicles are assumed to make left turns using jughandles, and automated vehicles making left turns must leave reserved lane and become a part of the general traffic in order to access jughandle ramp (Figure 1). Jughandle ramps on signalized corridors are frequently applied traffic regulation strategy and are well known to improve intersection capacity. This strategy along with reserved lanes for automated driving can further improve mobility performance of a signalized corridor.

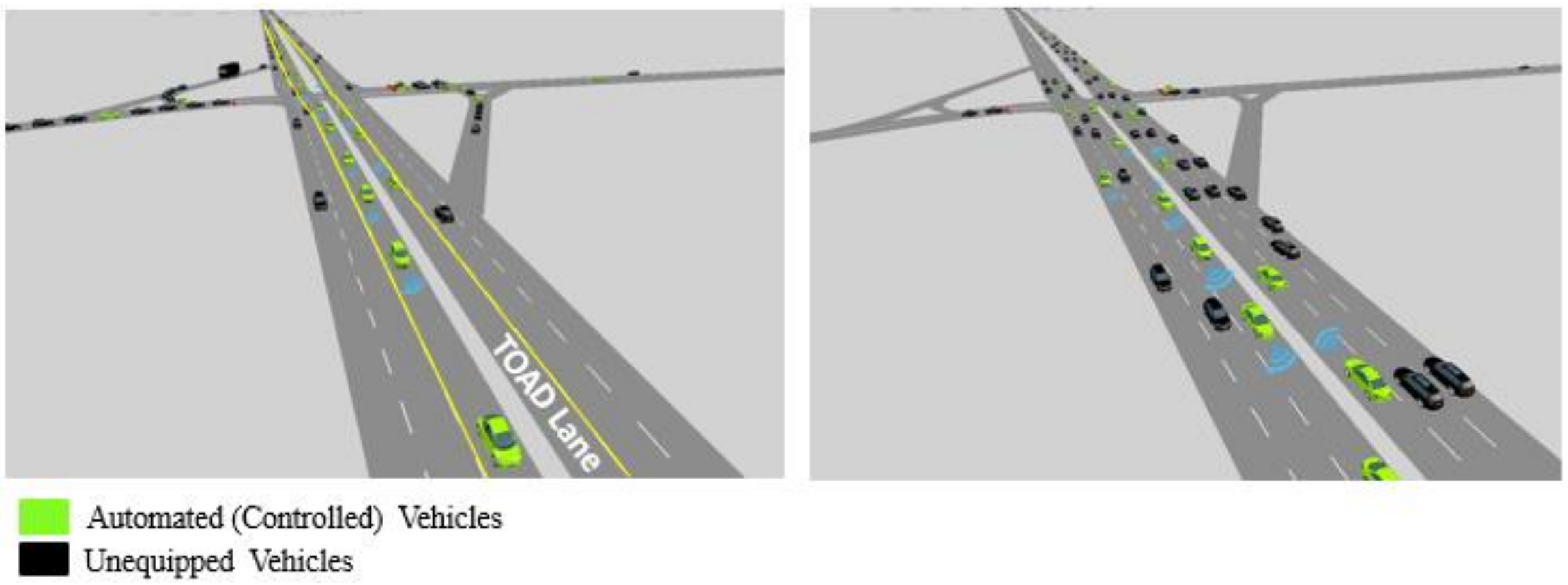

(a) With reserved lane for TOAD (b) Without reserved lane for TOAD

Figure 1. Simulation of the control strategy in PTV VISSIM.

**Test Bed 1: US-1 in Princeton, NJ**

The first test bed location (Figure 2) selected for the system evaluation is located in Princeton Township, New Jersey. A section of US-1 in Mercer County, between Carnegie Center Boulevard and Ridge Road, is about 5 miles long, with mainly six lanes in two directions. Coordinated intersections include jug-handle ramps with no left turns allowed from the mainline (i.e., US 1). The roadway has a speed limit of 55 mph. The corridor has numerous jughandle ramps allowing restriction of left turns from the mainline of the corridor.

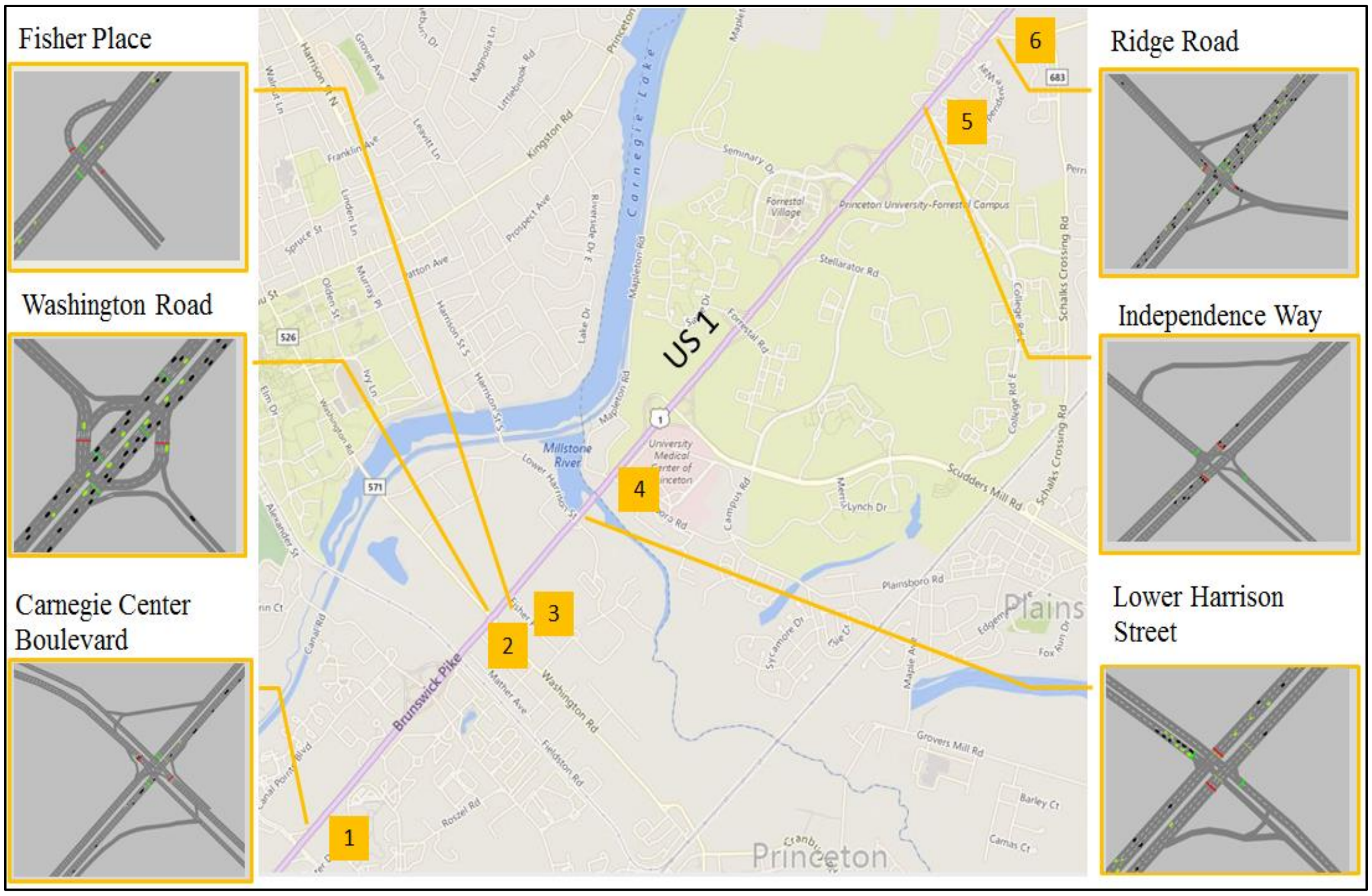

Figure 2. Test bed location in Princeton, NJ

**Test Bed 2: US-1 in Woodbridge, NJ**

Due to the different geometrical and traffic volume characteristics, the second testbed was developed to provide additional proof of the algorithm functionality. This corridor is characterized by different lengths of the corridor links, and slightly different volume rates in the observed time period. In addition, this corridor contains an isolated intersection and an intersection without a jughandle ramp. The second test corridor is presented in Figure 3. The test bed is located in Woodbridge Township, New Jersey. A section of US-1 in Middlesex County, between Gill Lane and Prince Street, is about 4 miles long, with mainly seven lanes in two directions (four lanes north-east, and three lanes south-west direction). Coordinated intersections include jughandle ramps with no left turns allowed from the mainline (i.e., US 1). The roadway has a speed limit of 55 mph.

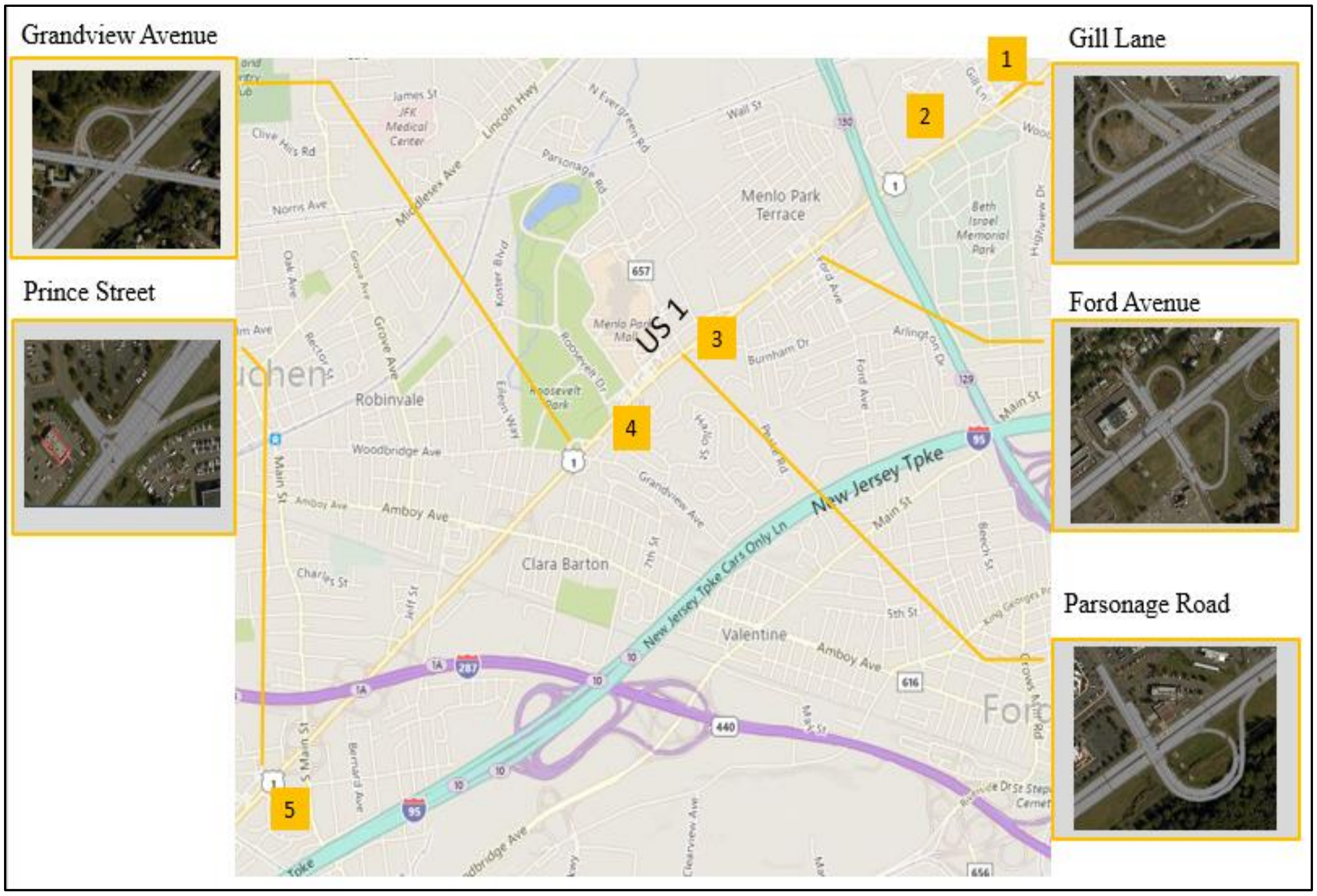

Figure 3. Tet bed location in Woodbridge, NJ

The simulation assessment comprises eleven different market penetration conditions (0%-100%, in 10% increments) that were examined with five consecutive simulation runs for cases with, and without reserved lanes for automated vehicles under the peak and off-peak traffic conditions. Simulation for eleven different market penetration conditions for both lane configurations was repeated five times, every time changing random seed parameter. Thus, this evaluation required a total of 110 simulation runs.

**Lane Configuration under Different Market Penetration and Volume Conditions**

Throughout simulation assessment, it was discovered that the best lane configuration setup depends on current traffic and market penetration levels. Specifically, under LOS A to C, the first reserved lane can be introduced as soon as market penetration level reaches 10% and additional reserved lane can be introduced with 50% of automated vehicles in the corridor. Under more congested conditions (i.e. LOS C to E) it is not recommended to assign any reserved lanes until the market penetration exceeded 60%. At this point, it is recommended to include two reserved lanes for automated vehicles. The reason for not including reserved lane before 60% market penetration can be explained using Figures 7 and 10. By observing those figures, it can be inferred that inclusion of reserved lanes before mentioned threshold might provide lower benefits compared to corridor without lane reservation.

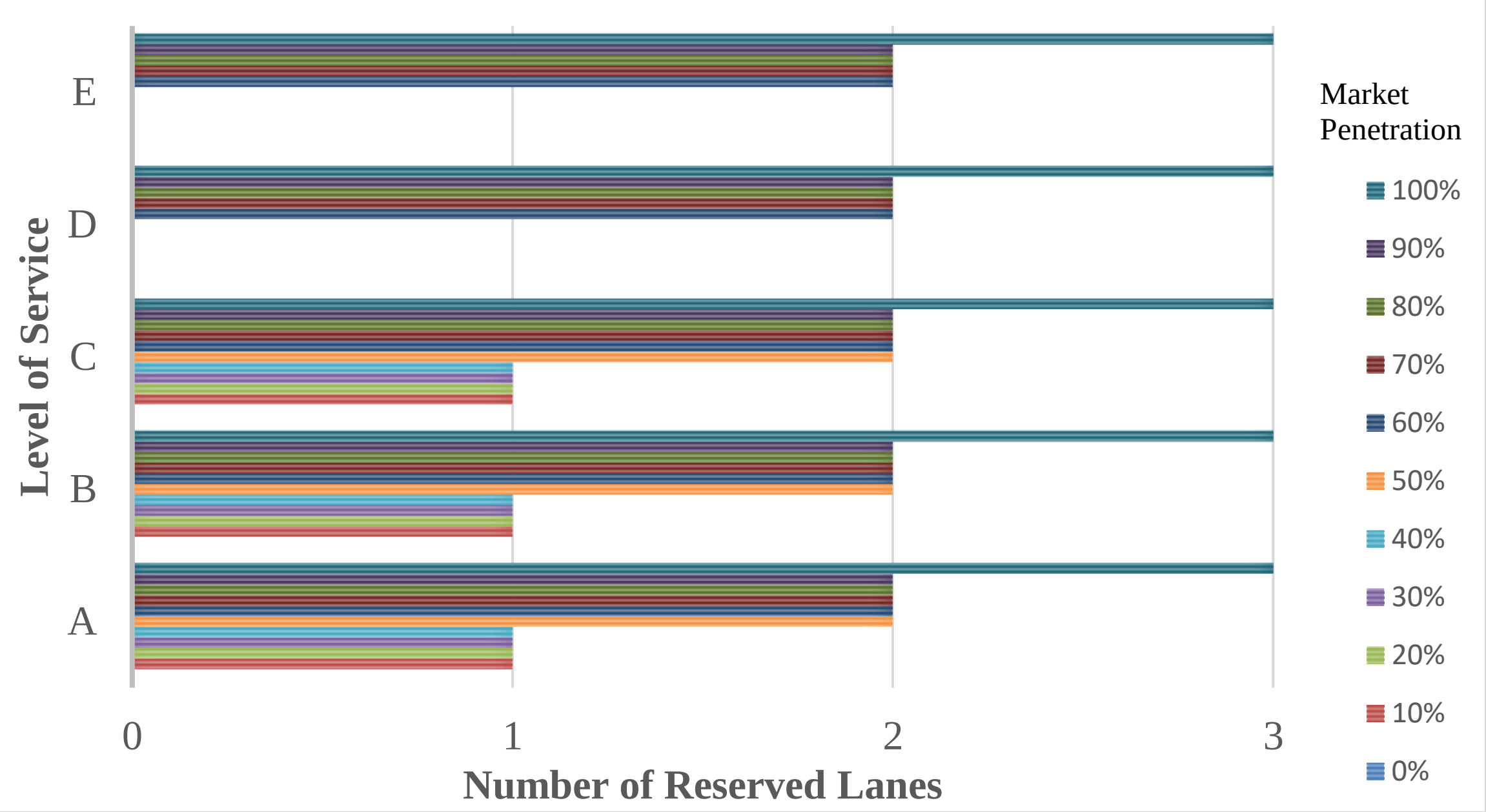

Figure 4. Recommended lane configuration for different traffic and market penetration levels

The main reason for lower benefits under unstable traffic conditions is an oversaturation of lanes assigned to general traffic under low market penetration levels. In such conditions, every time one or two lanes are assigned to automated vehicles, the lanes assigned to general traffic will be exposed to undesirable V/C ratios further influencing the average corridor travel time. The same conflict does not occur under lower LOS values, as the volume rates of the corridor are generally lower, making this limitation of the lane assignment strategy less visible.

**Evaluation Results for Testbed in Princeton, New Jersey**

It can be inferred from Figure 5, that overall trend of the average delay decreases with increased market penetration. The overall average delay is lower for a corridor with reserved lanes.

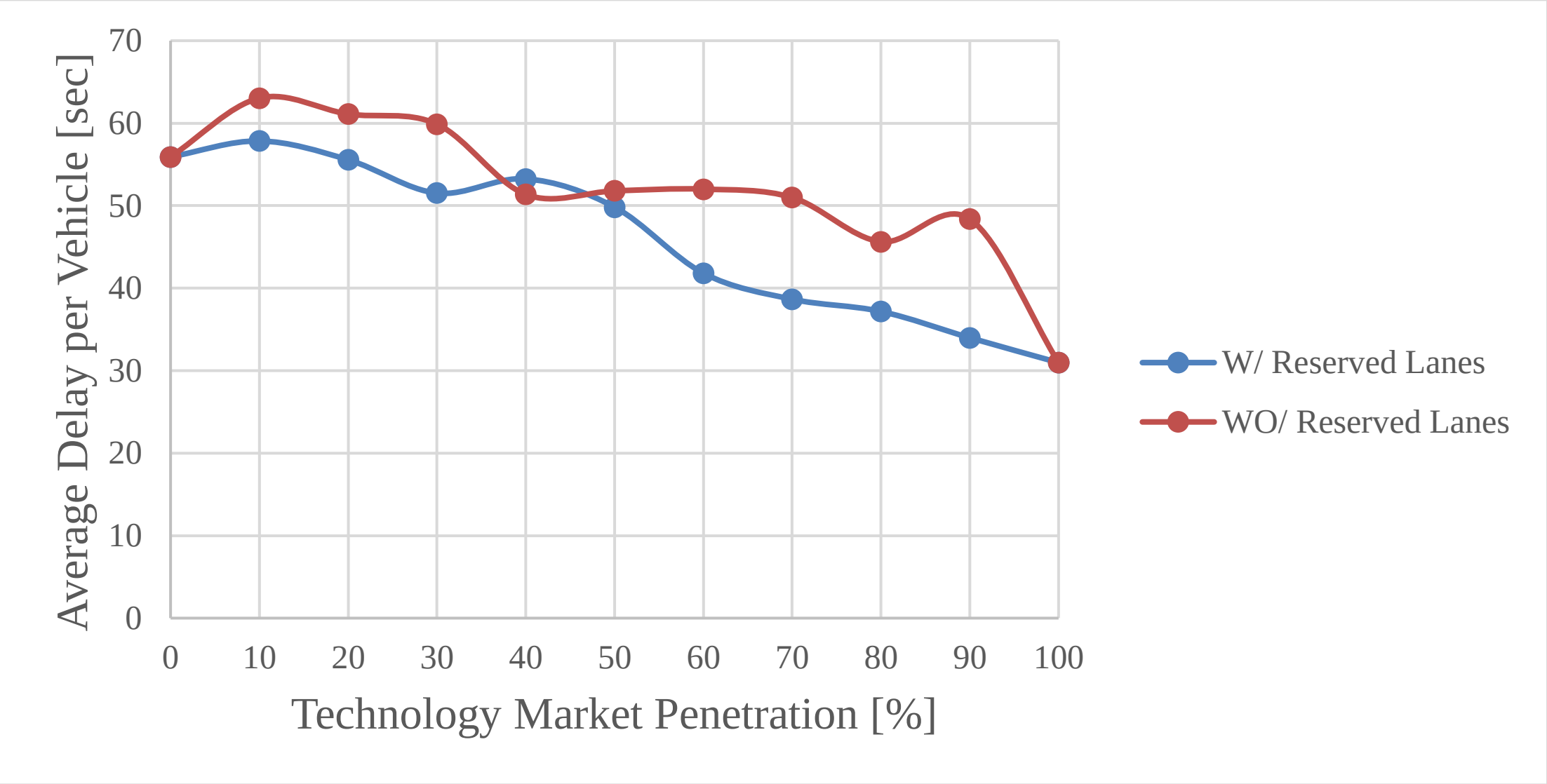

Figure 5. Average delay per vehicle for testbed in Princeton, NJ under LOS C.

The total number of served vehicles increased with a number of automated vehicles in the traffic stream (Figure 6). This trend was expected since decreased delay and a total number of stops provides better corridor progression. In addition, the more automated vehicles are present in the traffic stream, the less start-up lost time is experienced leading to increase in intersection throughputs along the corridor. Certain level of fluctuations was detected for the corridor with reserved lanes due to described change in a number of lanes reserved for such vehicles. The number of served vehicles is slightly higher for the corridor without reserved lanes. Since the overall utilization of such lanes fluctuates so does the overall throughput.

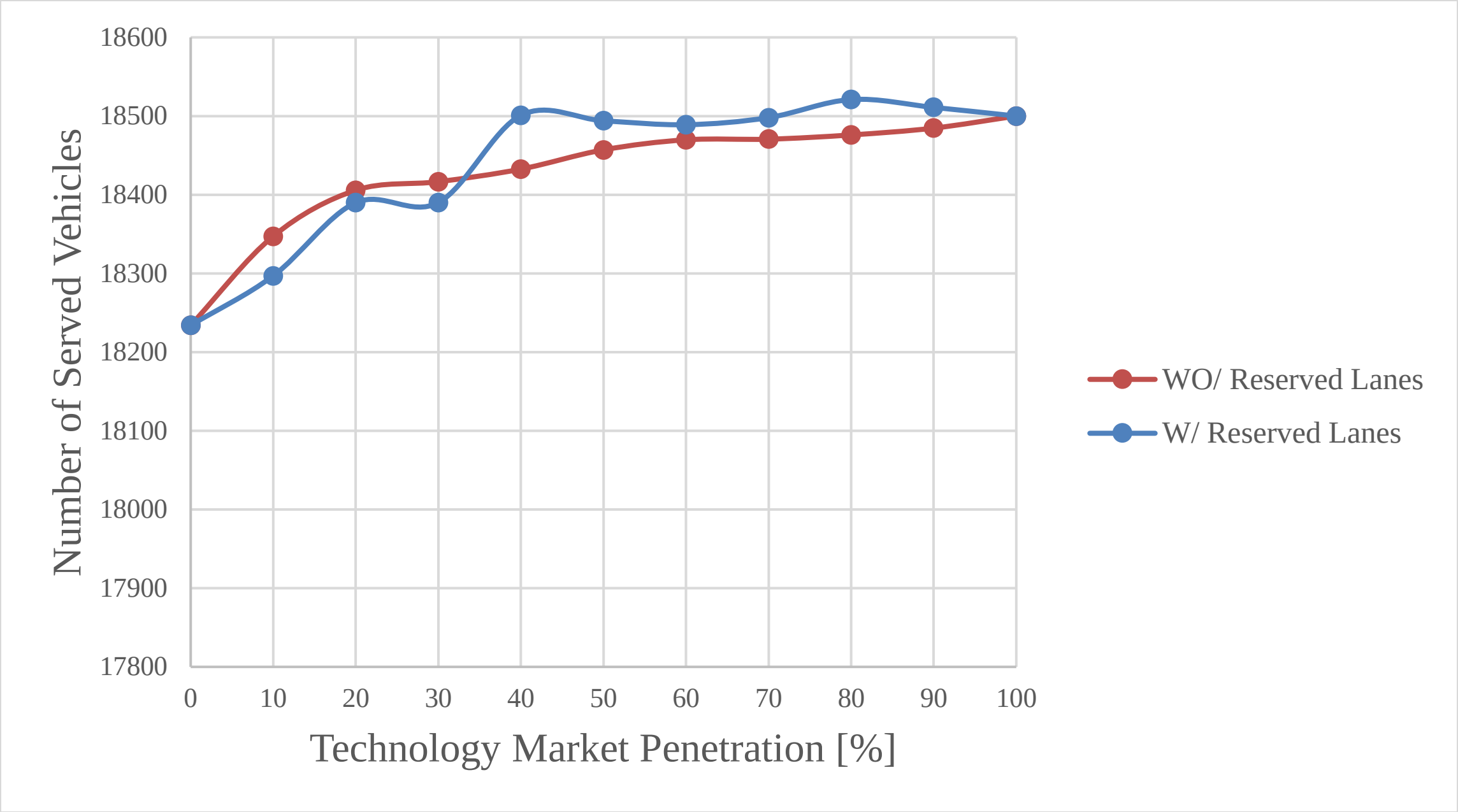

Figure 6. Total average number of served vehicles for testbed in Princeton, NJ, under LOS C.

The described mobility performance measures such as average vehicle delay, and increased throughput values are inevitably leading toward a decrease in overall travel time of the corridor. Nonetheless, the introduction of reserved lanes further increases the effectiveness of automated vehicles. Figure 7 illustrates the average total travel time for described corridor under different market penetration rates. Expectedly, the overall travel time decreases as a number of automated vehicles increases. Although benefits are marginal for low market penetration levels from 0-20% the overall functionality of the strategy is confirmed. Some more visible travel time reductions can be expected for market penetration rates higher than 20%. The trend of the travel time curves for the corridor with and without reserved lanes under given traffic conditions differs for market penetration levels from 20% to 80%. While the curve for the corridor without such lanes is almost flat for market penetrations 20-50%, the curve for the corridor with reserved lanes contains a certain level of decrease. Some more abrupt decrease is detected for market penetration rates higher than 50% in both lane configuration cases. For market penetration rates higher than 80%, the detected benefits are similar for both lane configuration cases as a result of a generally high number of automated vehicles in the traffic stream.

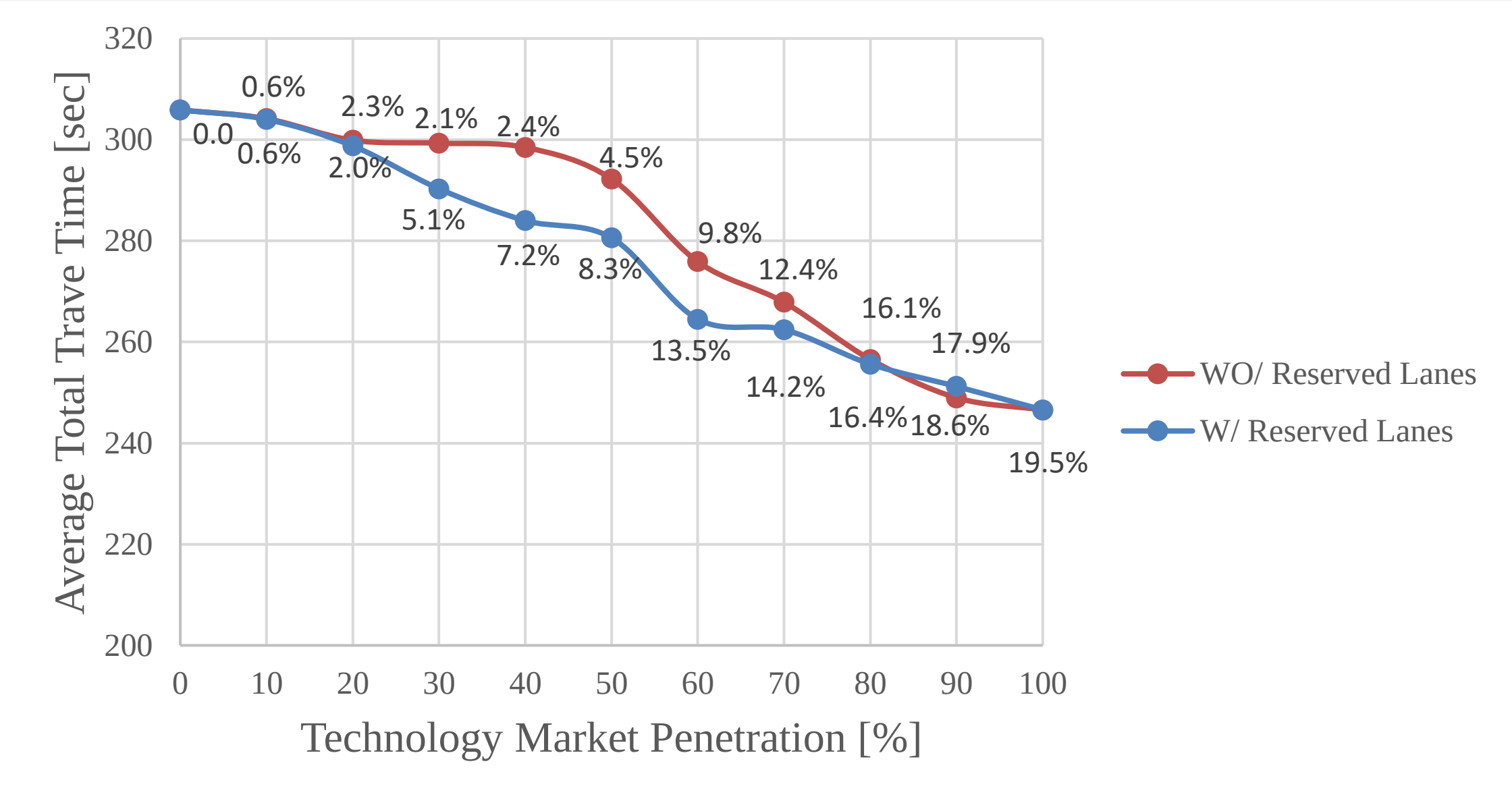

Figure 7. Total average travel time for testbed in Princeton, NJ under LOS C.

The total average delay reductions re less visible under LOS E. The total reductions under LOS C of 47% are now 13% due to congested corridor conditions. In such conditions, possibilities for generating more efficient vehicle trajectory are lower, as well as possibilities for lane changing. In general, delays are lower for the corridor with reserved lanes but the slope is significantly lower compared to the same curve generated under LOS C. The benefits for market penetration rates under 30% are almost similar, while after 30% they become more distinctive. An abrupt drop in average delay occurred after 80% of automated vehicles for the corridor without reserved lanes, while the overall trend of the same curve for the corridor with reserved lanes decreases gradually with increase in penetration level of automated vehicles.

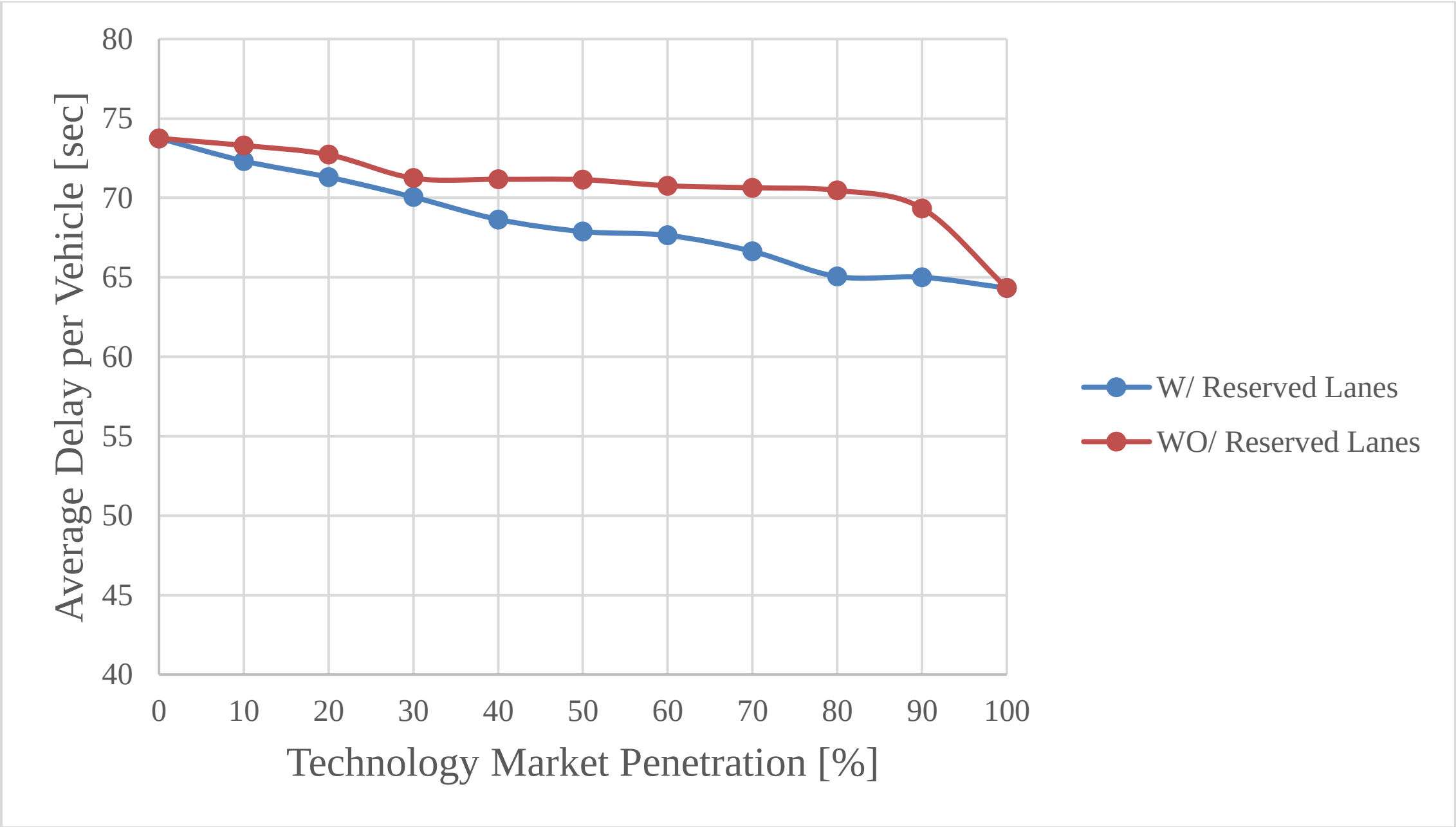

Figure 8. Total average delay under LOS E for testbed in Princeton, NJ.

Under LOS E the overall number of served vehicles increased by 8.5% (Figure 9). The LOS C revealed a slightly lower increase of 1.4% with 100% of automated vehicles in the traffic stream.

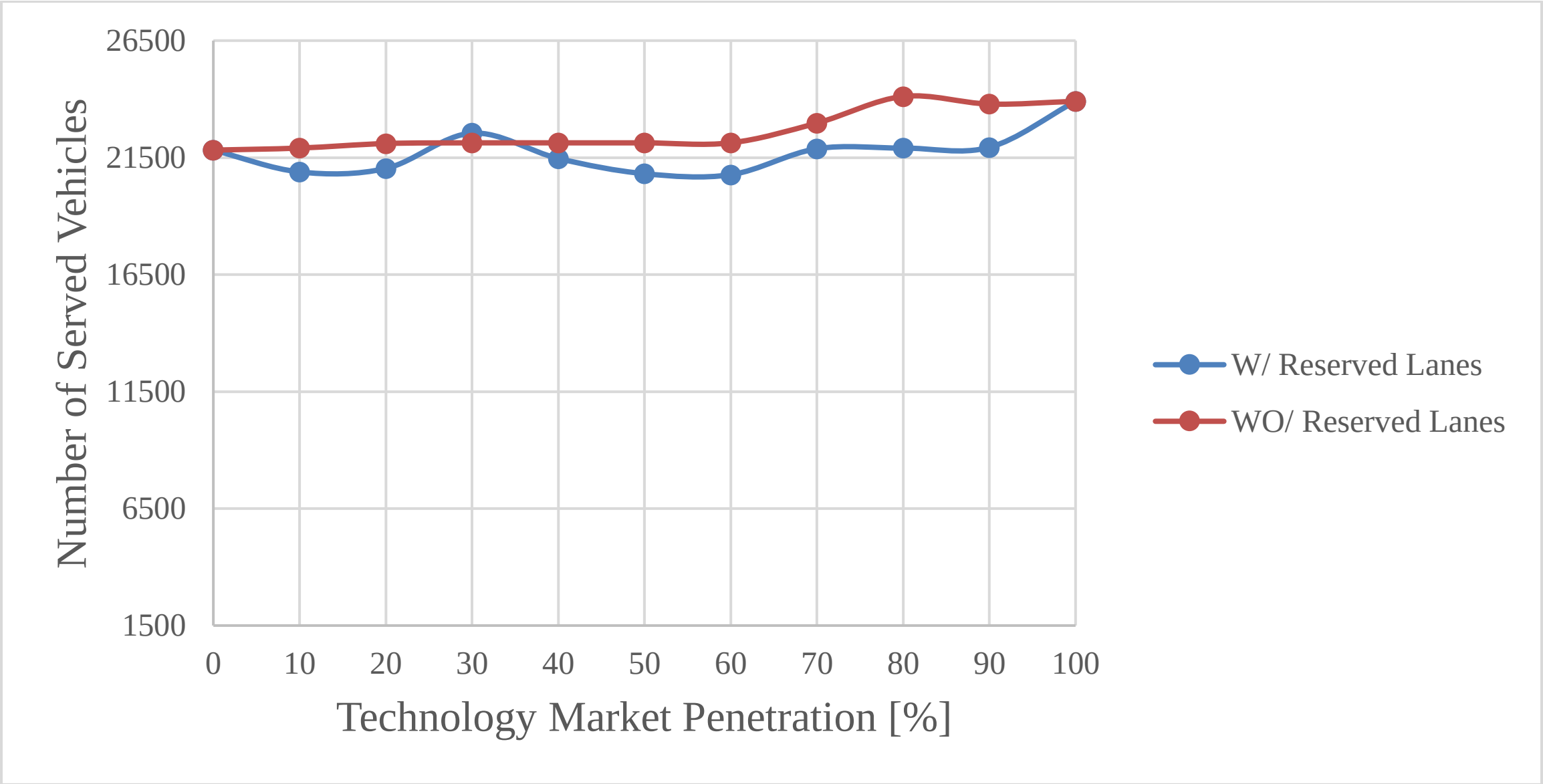

Figure 9. Total number of served vehicles under LOS E for testbed in Princeton, NJ.

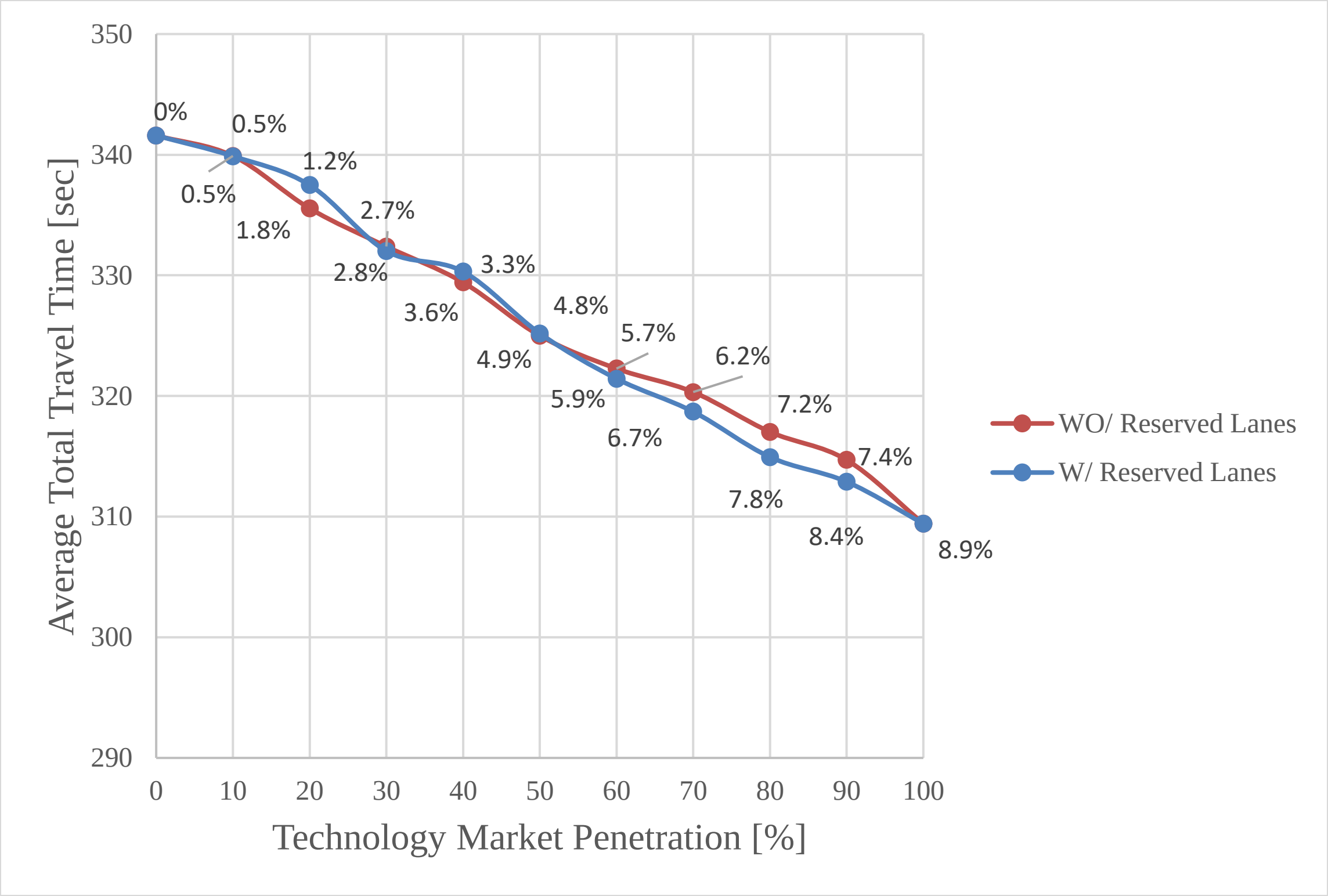

Figure 10. Average total travel time under LOS E for testbed in Princeton, NJ.

The congested corridor conditions not only decreased overall mobility improvements but also decreased the effectiveness of the reserved lanes. It can be inferred from Figure 10 that travel time curve for the corridor with and without reserved lanes achieved similar trends. Although the concept of reserved lanes is still more effective for market penetration higher than 60%, the difference in travel times is much more visible compared to uncongested conditions presented in the first part of this section. The main reason for the reduced effectiveness of the reserved lanes with respect to overall mobility performance lays in the fact that under congested traffic conditions general traffic suffers once a separate lane is assigned to automated vehicles due to reduced capacity. The same capacity reduction is less visible under uncongested traffic conditions as overall V/C ratios are lower allowing general traffic to operate with fewer constraints. With this finding, it is highly questionable if the concept of reserved lanes can be applied under highly congested traffic conditions.

**Evaluation Results for Testbed in Woodbridge, New Jersey**

The second testbed location generally confirmed findings outlined in the previous section. Maximal reductions in average delay under 100% market penetration are 45% and again have a decreasing trend as market penetration increases. The reductions are generally higher for the corridor with reserved lanes which confirmed findings from the first testbed location.

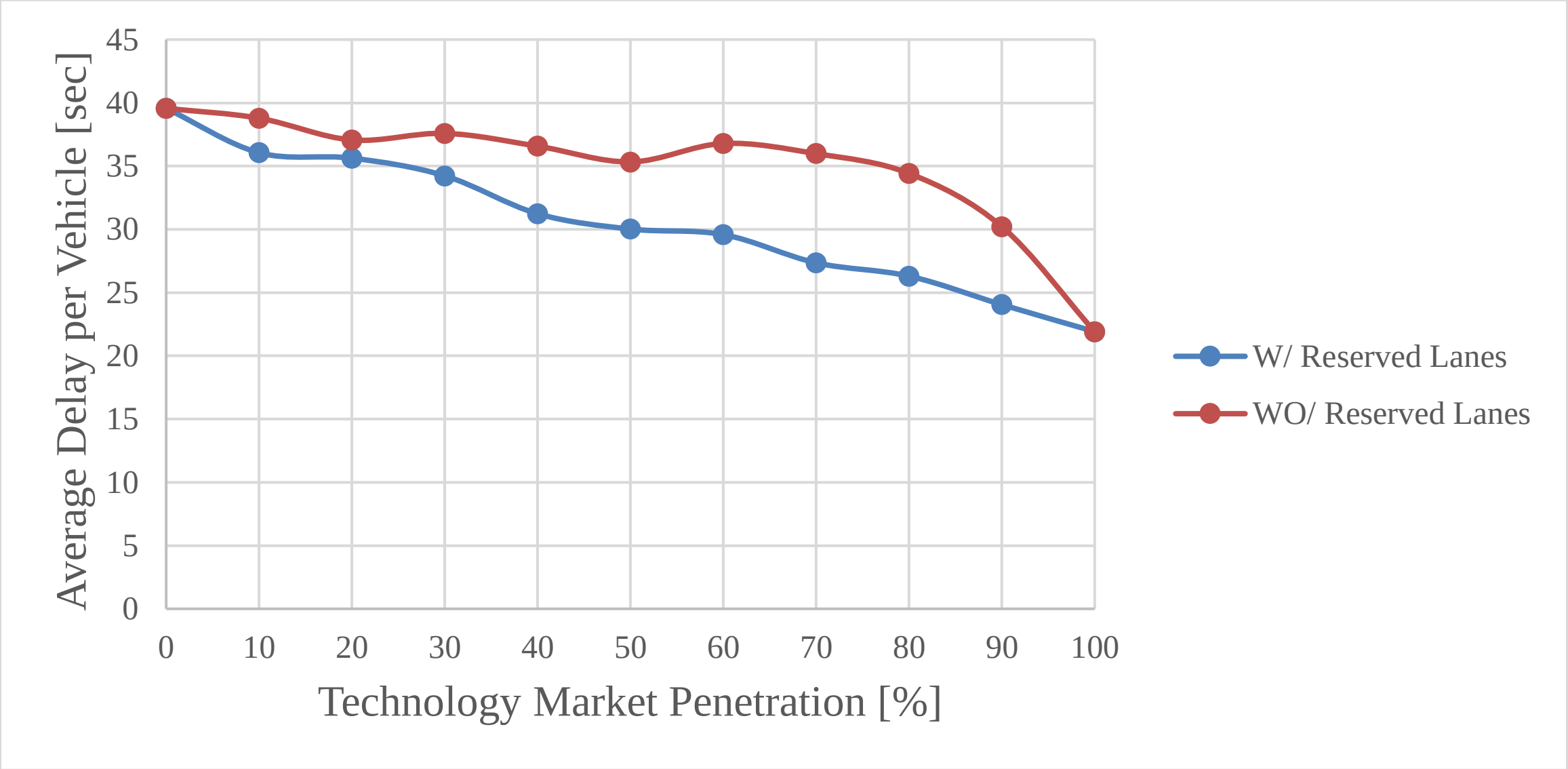

Figure 11. Total average delay for testbed in Woodbridge, NJ under LOS C

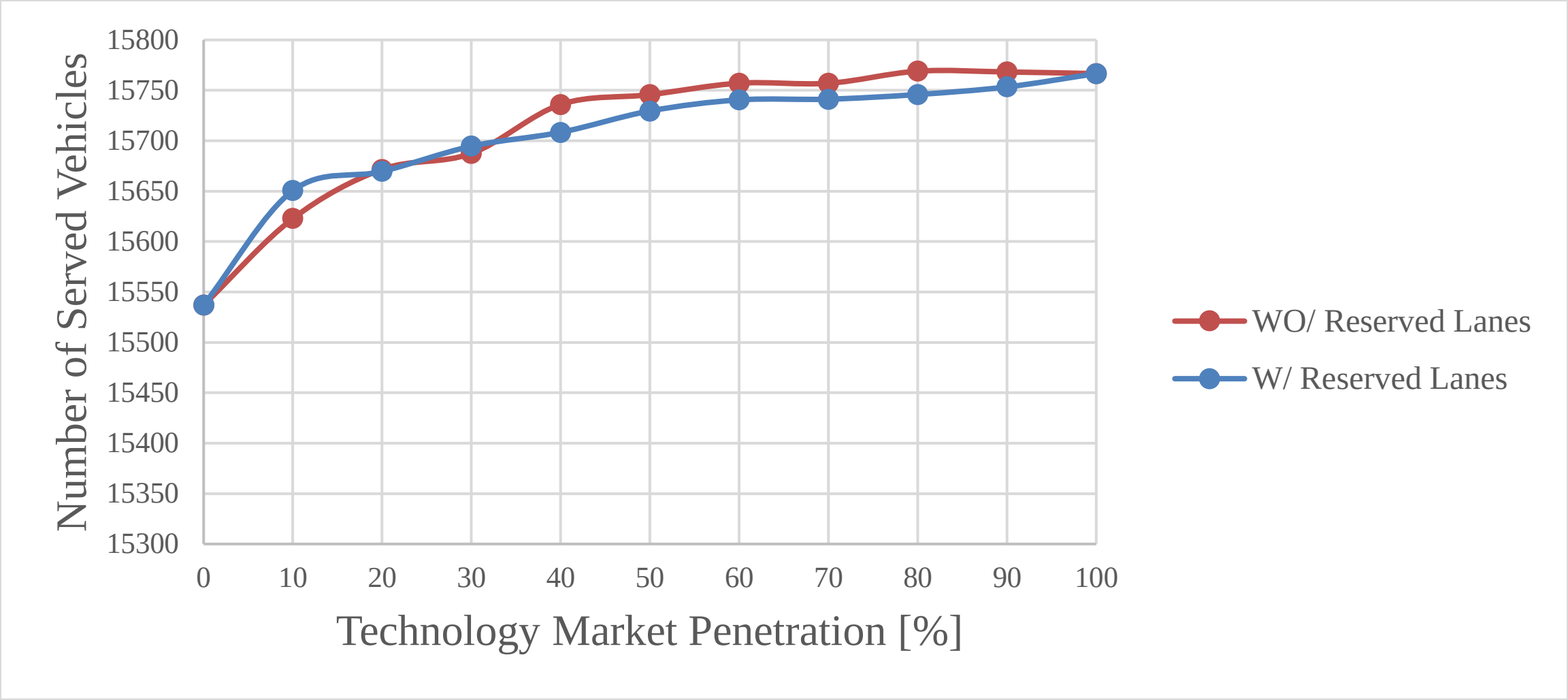

Figure 12. Total number of served vehicles for testbed in Woodbridge, NJ under LOS C.

The total number of the served vehicle increased by 8.7% under uncongested conditions. The increase in vehicles served for market penetration from 0-90% is higher with reserved lanes although both cases with and without reserved lanes revealed rising trend.

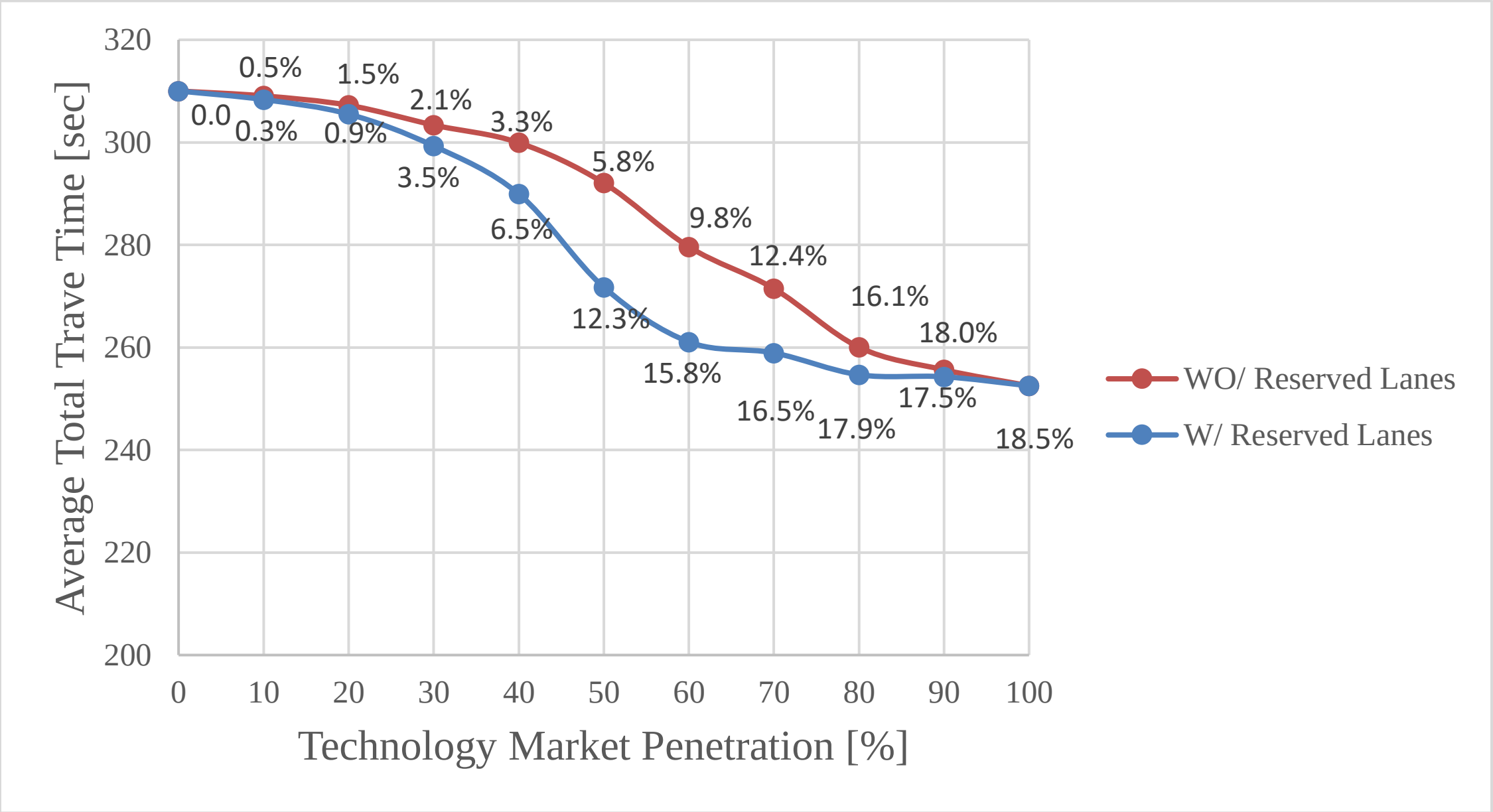

Figure 13**.** Average total travel time for testbed in Woodbridge, NJ under LOS C.

The lower market penetrations achieved marginal travel time reductions, however, uninterrupted corridor operation was achieved. Similar to previous testbed location, the benefits become more visible for market penetration rates higher than 20% of automated vehicles in the traffic stream. Again, the effect of reserved lane strategy is visible and provides a further increase of the algorithm effectiveness for market penetrations between 20% and 80%. The travel times for corridor with and without reserved lanes are similar once market penetration reaches 90%. Under 100% the entire corridor is occupied by equipped vehicles, so the travel time values are equal under 100% market penetration.

Further increase in corridor congestion level, revealed similar changes in observed parameters as those observed for the testbed in Princeton, New Jersey.

The average delay per vehicle increased due to an increased level of congestion on the corridor, but so did the overall reductions in average delay. Under LOS C such reductions were 72%. Under congested corridor conditions, such reductions dropped to 34% for 100% market penetration of automated vehicles.

The total average vehicle delays also increased under congested conditions. The magnitude of the average delay reductions also decreased. While under LOS C the maximal reduction under 100% market penetration was 45%, under congested conditions the reduction was 12%. The reductions are higher for the corridor with reserved lanes just as it was observed at the first testbed location in Woodbridge, New Jersey. The average delay values for different market penetration rates are illustrated in Figure 14.

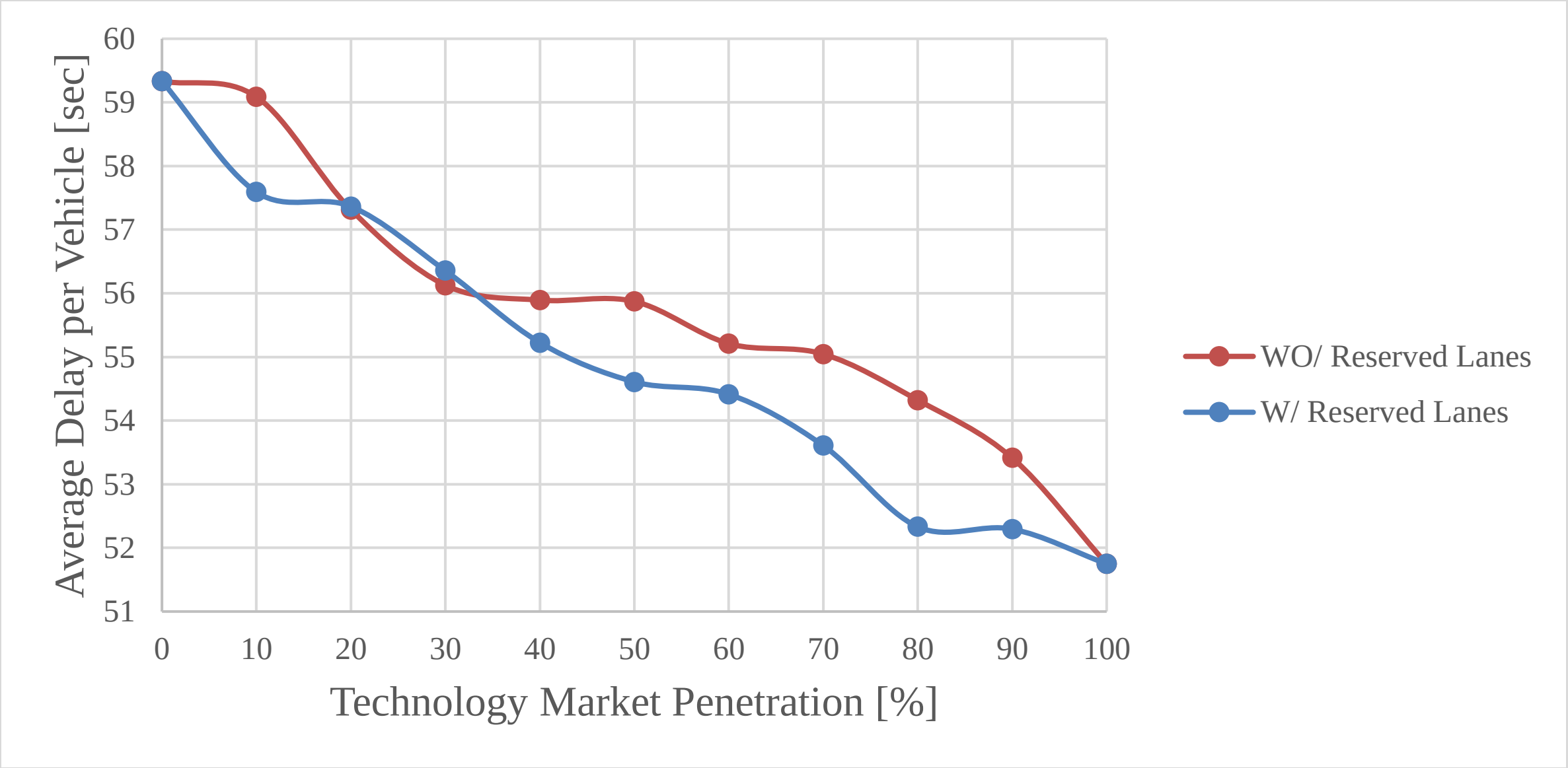

Figure 14. Total average delay for testbed in Woodbridge, NJ under LOS E.

The increase in throughput is visible under congested traffic conditions. Again, on the corridor with reserved lanes, the throughput is slightly lower and the maximal increase in throughput under 100% of automated vehicles is around 7%.

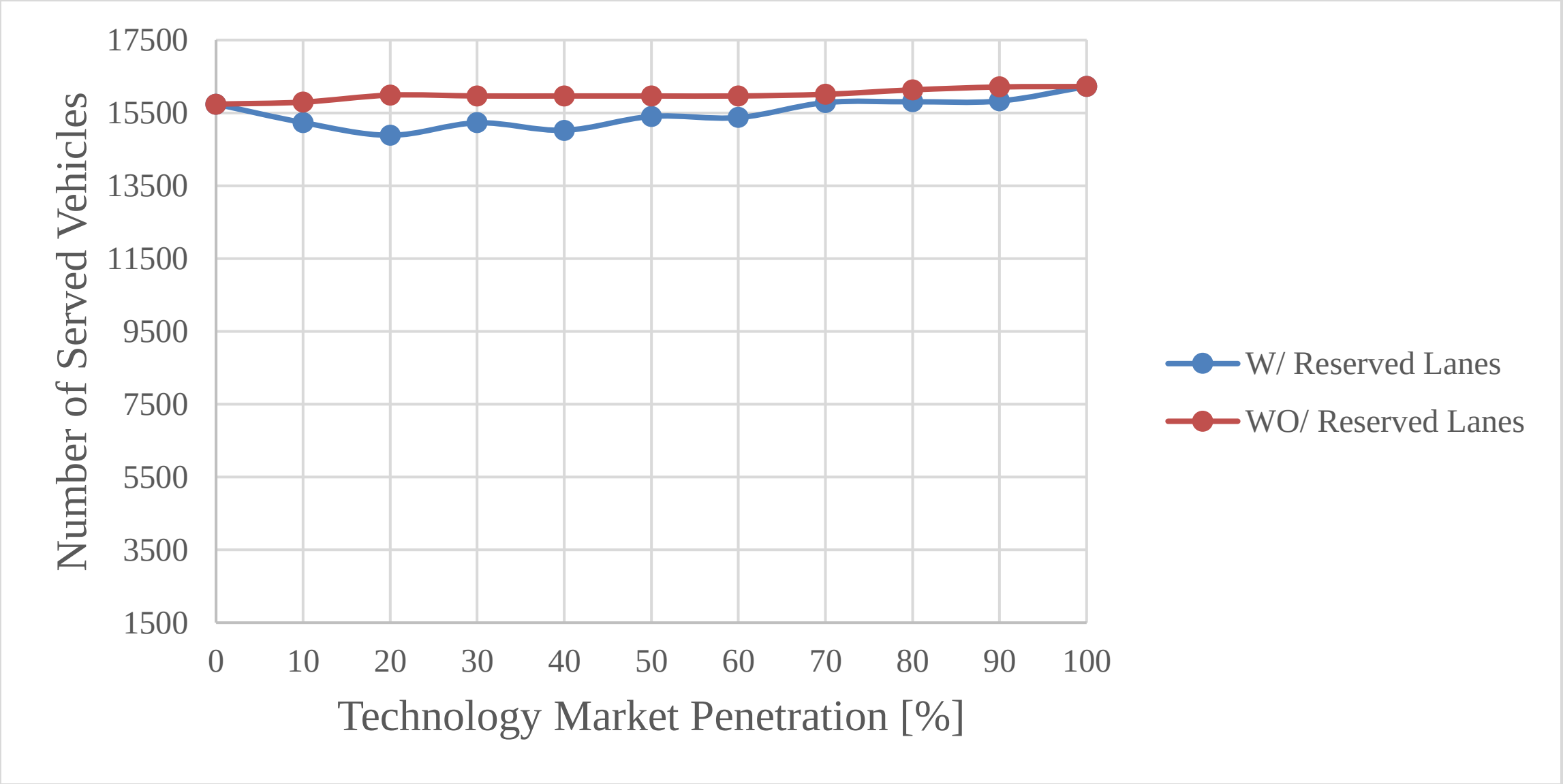

Figure 15. Total vehicles served for testbed in Woodbridge, NJ under LOS E.

Once again, the performance of the corridor with reserved lanes under congested traffic conditions on the signalized corridor in Woodbridge showed significantly lower benefits of reserved lanes compared to uncongested conditions. The findings with respect to corridor travel times correspond to those detected on the first testbed locations and are illustrated in Figure 16.

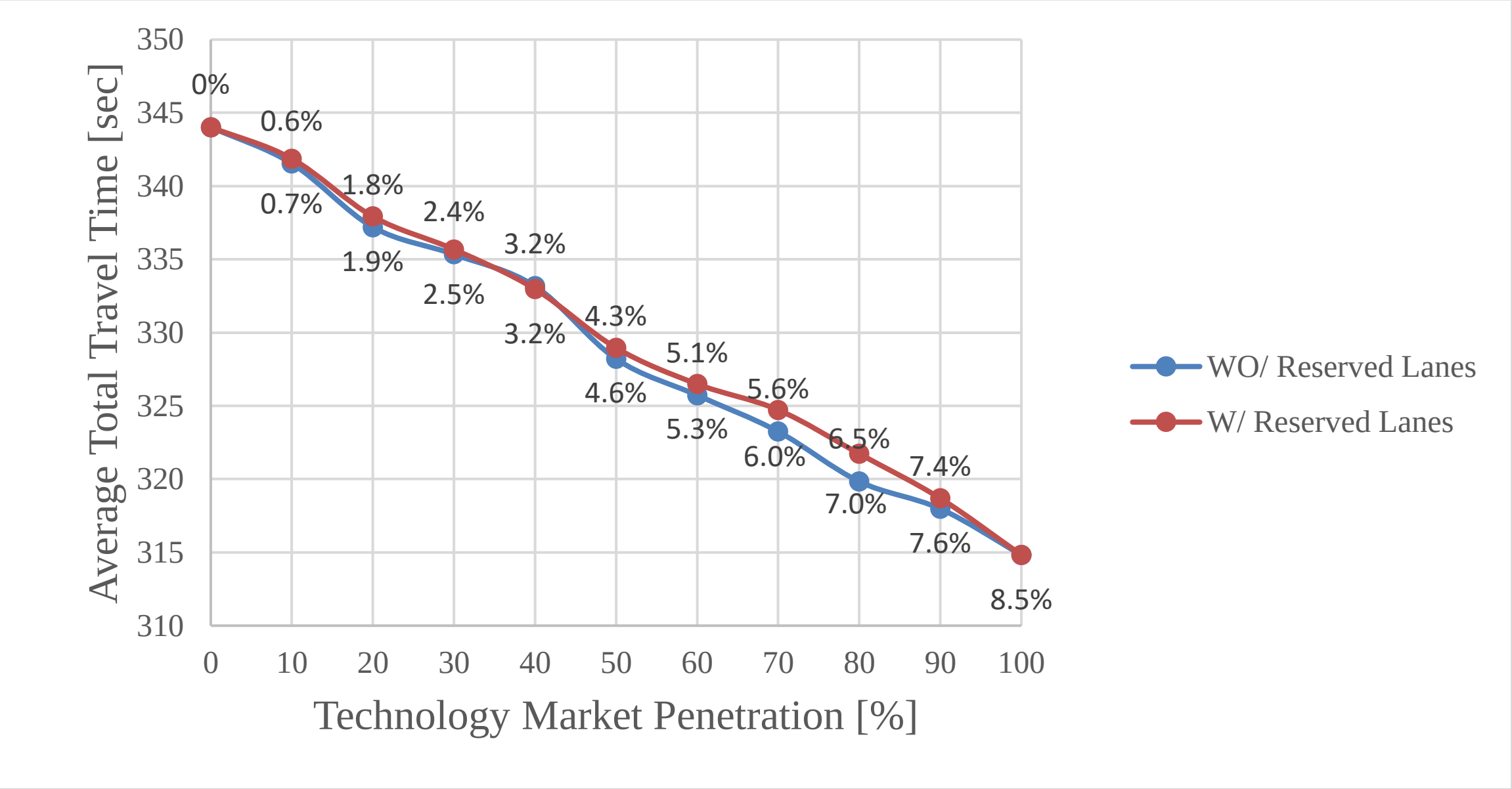

Figure 16. Average total travel time for testbed in Woodbridge, NJ under LOS E.

## CONCLUSION

This research investigated the impact of the lane reservation strategy on overall performance of the proposed corridor management. Under the low market penetration conditions, with less than 20% of automated vehicles in the stream, the inclusion of the reserved lane did not provide any improvements (0.5% to 0.6%). By further increasing the market penetration to 30% of automated vehicles in the corridor, the travel time savings of the corridor with no reserved lane for automated vehicles are lower (2.1% for the first and the second testbed location) than those of the corridor with reserved lanes (0.3.5 % - 5.1 % for the first and second testbed locations respectively). Under stable traffic conditions (i.e., LOS C), a further increase in market penetration levels brings additional benefits ranging from 2.4-19.4% and 3.3-18.5% for market penetrations of 40-100% for the first and second testbed location respectively, with no reserved lanes for automated vehicles.

The findings also imply that for the signalized corridors observed in this study, under given traffic conditions, the benchmark point for the introduction of a reserved lane is 30% of automated vehicles in the system. The simulation methodology also detected that the second reserved lane for automated vehicles should be included when the proportion of automated vehicles exceeds 50%. Under those lane configuration cases, total reductions in total corridor travel times are ranging from 5.1% to 19.4% and 6.5% to 18.5% for market penetrations of 40-100% for the first and second testbed location, respectively. Congested traffic conditions (i.e., LOS E) are also examined for both lane configuration cases. It was discovered that additional benefits produced by lane reservation strategy produced fewer improvements in such traffic conditions. Although all observed traffic stream parameters (average vehicle delay, number of served vehicles and total average travel time) generally decrease as market penetration decreases, the inclusion of reserved lanes brings marginal improvements for market penetrations above 60%. For the first test bed location the travel time reductions with reserved lanes is ranging from 5.9% to 7.4% (5.1% to 7.4% for the second testbed location) for market penetrations of 60-90% while the travel time reductions for the corridor without reserved lanes are ranging from 5.7% to 7.8% (5.3% to 7.6% for the second testbed location). For market penetration levels below 60%, both corridors showed insignificant differences between travel time results for cases with and without reserved lanes.

Total travel time reductions with 100% market penetration decrease from 19.5% to 8.5% under LOS C and LOS E respectively. Such reduced benefits under significantly congested conditions still provide noticeable benefits and allow an uninterrupted operation of automated vehicles under imperfect market penetration rates.